\documentclass[pra,nobibnotes,showpacs,preprintnumbers,amsmath,amssymb]{revtex4}

\usepackage{graphicx} 
\usepackage{bm}
\newcommand{\rd}{{\rm d}}

\newcommand{\kL}{k_{\rm L}}
\newcommand{\ER}{E_{\rm R}}
\newcommand{\vR}{v_{\rm R}}

\newcommand{\Name}[1]{#1,}%{\mu}
\newcommand{\REVIEW}[4]{{ #1} {\bf #2}, #4 (#3).}%{\mu}
\newcommand{\Book}[1]{{\it #1,}}
\newcommand{\Year}[1]{(#1).}
\newcommand{\Publ}[1]{#1,}

\begin{document}
\title{Bogoliubov speed of sound for a dilute Bose--Einstein condensate
in a 3d optical lattice}

\author{Dave Boers}
\author{Christoph Weiss}
\author{Martin Holthaus} 
\affiliation{Institut f\"ur Physik, Carl von Ossietzky Universit\"at,
        D-26111 Oldenburg, Germany} 
%{Boson systems} % % % %******************** Abbreviations for maths ******************** 

\begin{abstract} 
We point out that the velocity of propagation of sound wavepackets in  
a Bose--Einstein condensate filling a three-dimensional cubic optical  
lattice undergoes a maximum with increasing lattice depth. For a realistic 
choice of parameters, the maximum sound velocity in a lattice condensate  
can exceed the sound velocity in a homogeneous condensate with the same  
average density by $30\%$. The maximum falls into the superfluid regime,  
and should be observable under currently achievable laboratory conditions. 
\end{abstract} 
 
\maketitle 
 
There have been vigorous activities, and impressive achievements, concerning  
Bose--Einstein condensates in optical lattice potentials recently, both  
experimentally~\cite{AndersonKasevich98,CataliottiEtAl01,MorschEtAl01,
        GreinerEtAl01,GreinerEtAl02a,GreinerEtAl02b} and  
theoretically~\cite{SorensenMolmer98,JakschEtAl98,ChoiNiu99,Javanainen99,
        ChiofaloEtAl00,KraemerEtAl02,MachholmEtAl03,SmerziTrombettoni03,
        KrutitskyGraham03},  
culminating in the observation of the quantum phase transition from  
a superfluid to a Mott insulator with a condensate of $^{87}$Rb  
atoms~\cite{GreinerEtAl02a}.   
 
On the theoretical side, substantial effort has been devoted to  
calculating the low-lying excitations of a condensate in an optical  
lattice, and thus the speed of sound, within the Bogoliubov  
theory~\cite{vanOostenEtAl01,ReyEtAl03,TaylorZaremba03}. In particular, it  
has been predicted that in a one-dimensional optical lattice the velocity of  
propagation of sound wavepackets decreases monotonically with increasing  
lattice depth~\cite{KraemerEtAl03,MenottiEtAl04}. In this letter, we argue  
that the situation is different for condensates in three-dimensional lattices:  
In this case the velocity of sound undergoes a pronounced maximum when the  
lattice is made successively deeper, which falls into the superfluid regime, 
and should be observable under currently achievable experimental conditions. 
 
We consider a 3d~Bose--Einstein condensate subjected to a simple  
$d$-dimensional optical cosine lattice ($d = 1,3$) of depth~$V_0$,  
\begin{equation} 
        V({\bm r}) = \frac{V_0}{2} \sum_{i=1}^d \cos(2\kL x_i) \; ,  
\label{eq:CPot} 
\end{equation} 
where $\kL = 2\pi/\lambda$ denotes the wavenumber associated with the  
lattice-generating laser radiation of wavelength~$\lambda$. For $d=1$ 
we study the propagation of sound in the direction of the lattice, while  
assuming that the condensate remains homogeneous in the orthogonal plane.  
We also assume that the depth~$V_0$ of the lattice be sufficiently deep so  
that only the lowest Bloch band needs to be taken into account.   
 
The velocity of sound propagation in a Bose--Einstein condensate filling 
the lattice then is given by the standard expression~\cite{KraemerEtAl03, 
        MenottiEtAl04} 
\begin{equation} 
        c = \sqrt{\frac{U}{m^*}} \; , 
\label{eq:Vels} 
\end{equation} 
where $m^*$ is the effective mass pertaining to the lowest band, 
and~$U$ is the inverse compressibility of the condensate in the lattice. 
 
In principle, also the effective mass~$m^*$ appearing in eq.~(\ref{eq:Vels}) 
does depend on the density of the condensate, and has to be determined 
by solving the Gross--Pitaevskii equation for the condensate in the  
lattice~\cite{SorensenMolmer98,ChiofaloEtAl00,MachholmEtAl03,TaylorZaremba03}.  
This is particularly pertinent if one considers a 1d~lattice filled with an  
effectively 1d~condensate, strongly confined by restoring potentials orthogonal  
to the lattice direction, so that the transversal degrees of freedom are 
frozen out entirely. If then each lattice site carries a comparatively large  
number of particles, nonlinear effects due to the mean-field interaction play  
a decisive role. In contrast, we focus here on truly 3d~condensates,  
with an average density $N/V$ of presently achievable magnitude. Taking the  
moderate value $N/V = 10^{13}$~cm$^{-1}$, say, and a lattice constant  
$\lambda/2 = 426$~nm, one obtains an average occupancy of about $0.8$ atoms 
per unit cell in a 3d lattice. Thus, we define the site occupancy~$N_s$ 
by writing 
\begin{equation} 
        \frac{N}{V} = \frac{N_s}{(\lambda/2)^3} \; , 
\label{eq:Dens} 
\end{equation} 
and consider values of $N_s$ which are on the order of unity. In this case,  
the density dependence of~$m^*$ is negligible, as wittnessed by numerical  
calculations for $d = 1$~\cite{KraemerEtAl03,MenottiEtAl04}, so that we  
obtain quite accurate predictions by invoking the effective mass, and  
the band structure, of the single-particle problem. 
 
Moreover, for low site occupancies it is not necessary to solve the  
Bogoliubov--de Gennes equations in order to obtain the elementary 
excitations, but we may perform the Bogoliubov transformation directly 
on the basis of the single-particle Bloch waves~$u_{\bm k}({\bm r})$.  
We then find for the inverse compressibility the convenient expression  
\begin{equation} 
        U = U_0  \; \frac{1}{\Omega} 
        \int_{\rm unit~cell} \! \rd^d {\bm r} \, | u_{\bm 0}({\bm r}) |^4 \; ,  
\label{eq:Rena}  
\end{equation} 
where $\Omega = (\lambda/2)^d$ is the volume of the unit cell, and 
$u_{\bm 0}({\bm r})$ is the Bloch function corresponding to zero quasimomentum, 
normalized such that $\int \! \rd^d {\bm r}\, | u_{\bm 0}(\bm r) |^2 = \Omega$, 
with the integral again extending over one unit cell. The factor  
\begin{equation} 
        U_0 = \frac{4\pi a\hbar^2}{m} \frac{N}{V} 
\label{eq:Uzer} 
\end{equation} 
coincides with the customary interaction energy parameter characterizing 
a homogeneous Bose--Einstein condensate with a density of~$N$ particles 
per volume~$V$, which consists of atoms with (bare) mass~$m$ possessing 
the $s$-wave scattering length~$a$.  
 
One can intuitively capture the physics expressed by eq.~(\ref{eq:Rena})  
by imagining an initially homogeneous 3d~condensate within which a  
$d$-dimensional optical lattice is errected adiabatically; the particles  
then gradually get concentrated around the local minima of the lattice  
potential. This is described by the fact that the Bloch function  
$u_{\bm 0}(\bm r)$ develops a maximum within each well; the integral over the  
fourth power of this function, taken over one unit cell, increases slowly with  
increasing lattice depth. Obviously, the density reached at the potential  
minima becomes the higher, the larger the lattice dimension~$d$; as a  
consequence, the compressibility~$U^{-1}$ of the gas within the lattice also  
exhibits a significant $d$-dependence. It is this feature which translates  
itself into a dependence of the velocity of sound on the lattice depth which  
is markedly $d$-dependent.  
     
Before presenting numerical data for the velocity of sound in one- and 
three-dimensional lattices, it is useful to consider the analytically 
tractable limiting case of a deep optical lattice. Introducing the  
single-photon recoil energy~$\ER = \hbar^2 \kL^2 / (2m)$, and the associated  
recoil velocity $\vR = \hbar \kL / m$, we rewrite eq.~(\ref{eq:Vels}) in  
terms of convenient dimensionless ratios, obtaining  
\begin{equation} 
        \frac{c}{\vR} = \sqrt{\frac{U}{2\ER} \, \frac{m}{m^*}} \; . 
\label{eq:Vrat} 
\end{equation} 
Expanding the lattice potential quadratically around the minima of the  
cosine wells, the Bloch function $u_{\bm 0}(\bm r)$ can be approximated by 
a superposition of the groundstate wave functions of the corresponding  
harmonic oscillators, if the lattice is sufficiently deep. This harmonic 
approximation immediately leads to the estimate  
\begin{equation}  
        \frac{1}{\Omega}    
        \int_{\rm unit~cell} \! \rd^d {\bm r} \, | u_{\bm 0}({\bm r}) |^4 
        \sim \left( \frac{\pi^2}{4} \frac{V_0}{\ER}\right)^{d/4}  
        \qquad \mbox{for} \; V_0/\ER \gg1 \; , 
\label{eq:EstU} 
\end{equation}   
where here and in the following the ``$\sim$''-sign means asymptotic 
equality. By comparison with exact numerical data, we infer that for $d = 1$  
and $V_0/\ER = 10$ this approximation is about 12\% too high; the error  
decreases to 5\% when $V_0/\ER = 25$.  
 
This estimate now allows us to specify the condition for the validity 
of our approach more precisely: Since the gap between the lowest two  
single-particle Bloch bands amounts to $V_0/2$, and that gap has to be large  
in comparison with the interaction energy per particle in order to justify  
the restriction to the lowest band, we require $V_0 \gg U$. Using the  
above approximation in eq.~(\ref{eq:Rena}) for~$U$, together with the  
expression~(\ref{eq:Dens}) for the density, this gives for $d = 3$ the  
condition 
\begin{equation} 
        \left(\frac{V_0}{\ER}\right)^{1/4} \gg 
        4 \sqrt{2\pi} \, N_s \frac{a}{\lambda}  \; . 
\end{equation} 
Since the ratio $a/\lambda$ of the $s$-wave scattering length~$a$ to 
the laser wavelength $\lambda$ usually is of the order of $1/100$, 
this condition indeed limits the validity of our reasoning to occupancies 
$N_s$ of a few atoms per site, a ``site'' corresponding to the volume 
$(\lambda/2)^3$.  
 
The theory of the Mathieu equation now allows one to state an approximate 
expression for the width~$\Delta$ of the lowest energy band when the cosine  
lattice is sufficiently deep, namely~\cite{AbramowitzStegun72,Zwerger03} 
\begin{equation} 
        \frac{\Delta}{\ER} \sim \frac{16}{\sqrt{\pi}} 
        \left(\frac{V_0}{\ER}\right)^{3/4}  
        \exp\!\left(-2\sqrt{V_0/\ER}\right) \; . 
\label{eq:EstW} 
\end{equation} 
For $V_0/\ER = 10$, this estimate still is about 18.5\% too high; the error  
becomes less than 5\% only if $V_0/\ER > 90$. Since a quadratic expansion  
of the tight-binding cosine energy dispersion relation readily connects  
the band width with the effective mass, 
\begin{equation} 
        \frac{m^*}{m} = \frac{4}{\pi^2} \frac{1}{\Delta/\ER} \; , 
\end{equation} 
we obtain 
\begin{equation} 
        \frac{m^*}{m} \sim \frac{1}{4\pi^{3/2}} 
        \left(\frac{V_0}{\ER}\right)^{-3/4}  
        \exp\!\left(2\sqrt{V_0/\ER}\right)  
        \qquad \mbox{for} \; V_0/\ER \gg1 \; . 
\label{eq:EstM} 
\end{equation} 
 
Combining now the deep-lattice estimates~(\ref{eq:EstU}) and (\ref{eq:EstM}), 
the velocity of sound~(\ref{eq:Vrat}) takes the form    
\begin{equation} 
        \frac{c^{d=1}}{\vR} \sim 
        2^{1/4} \pi \sqrt{\frac{U_0}{\ER}} \,  
        \left(\frac{V_0}{\ER}\right)^{1/2} 
        \exp\!\left(-\sqrt{V_0/\ER}\right) 
\label{eq:AS1} 
\end{equation} 
for $d = 1$, while for $d = 3$ we find  
\begin{equation} 
        \frac{c^{d=3}}{\vR} \sim 
        \frac{\pi^{3/2}}{2^{1/4}} \sqrt{\frac{U_0}{\ER}} \, 
        \left(\frac{V_0}{\ER}\right)^{3/4} 
        \exp\!\left(-\sqrt{V_0/\ER}\right) \; , 
\label{eq:AS3} 
\end{equation} 
assuming $V_0/\ER \gg 1$ in both cases. The exponential decrease of these 
velocities with increasing lattice depth is due to the increase of the 
effective mass~(\ref{eq:EstM}), or, equivalently, the reduction of the  
tunneling contact between the wells, while the different exponents appearing  
in the prefactors can be traced to the $d$-dependent inverse  
compressibilities~(\ref{eq:Rena}). 
 
In needs to be kept in mind, however, that when the lattice depth $V_0$  
exceeds a critical magnitude, the transition from the superfluid to the  
Mott insulator state  
occurs~\cite{GreinerEtAl02a,JakschEtAl98,FisherEtAl89,Zwerger03}.  
Since the Bogoliubov theory treats the interaction only approximately, it is  
incapable of describing large depletions of the condensate, and thus does not 
incorporate the transition~\cite{vanOostenEtAl01}. Hence, the Bogoliubov speed  
of sound~(\ref{eq:Vels}), and the above estimates, are meaningful only in the  
superfluid regime, sufficiently remote from the transition point. Within the  
tight-binding approximation, and assuming unit occupancy $N_s = 1$, in a  
three-dimensional lattice the transition takes place if the ratio of the  
on-site interaction energy per particle, 
\begin{equation} 
        \widetilde{U} = \frac{4\pi a \hbar^2}{m}  
        \int \! {\rm d}^3{\bm r} \, | w({\bm r}) |^4 \; , 
\label{eq:Inte} 
\end{equation} 
and the hopping matrix element $\Delta/4$ adopts the critical value 
\begin{equation} 
        \left(\frac{4 \, \widetilde{U}}{\Delta}\right)_{\! c}  
        \approx z \times 5.8 \; , 
\label{eq:Crit} 
\end{equation} 
where $z = 6$ is the number of nearest neighbours of each site. (For high 
$N_s$, the critical ratio approaches $4 N_s z$~\cite{Zwerger03}.) The function 
$w({\bm r})$ appearing in eq.~(\ref{eq:Inte}) is the Wannier function for 
the lowest band. Approximating this Wannier function once again by the  
groundstate wave function of the harmonic oscillator corresponding to 
a quadratically approximated cosine well~\cite{Slater52}, we find 
\begin{equation} 
        \frac{\widetilde{U}}{\ER} \sim  
        4 \sqrt{2\pi} \, \frac{a}{\lambda}  
        \left(\frac{V_0}{\ER} \right)^{3/4}     \; . 
\end{equation} 
In conjunction with the approximate relation~(\ref{eq:EstW}) for the band  
width, the criterion~(\ref{eq:Crit}) then allows us to estimate the critical  
lattice depth for $d = 3$ as    
\begin{equation} 
        \frac{V_{\rm c}}{\ER} \approx  
        \frac{1}{4} \ln^2 \left( 7.83 \frac{\lambda}{a} \right) \; . 
\end{equation} 
Actually, this simple estimate appears to be quite reliable: Considering 
$^{87}$Rb atoms, one has an $s$ wave scattering length of $a = 6$~nm.  
Choosing $\lambda = 852$~nm then gives $V_c/\ER \approx 12.3$, in perfect  
agreement with what has been observed in the experiment~\cite{GreinerEtAl02a}.  
 
\begin{figure} 
\includegraphics[width=0.8\linewidth,angle=0]{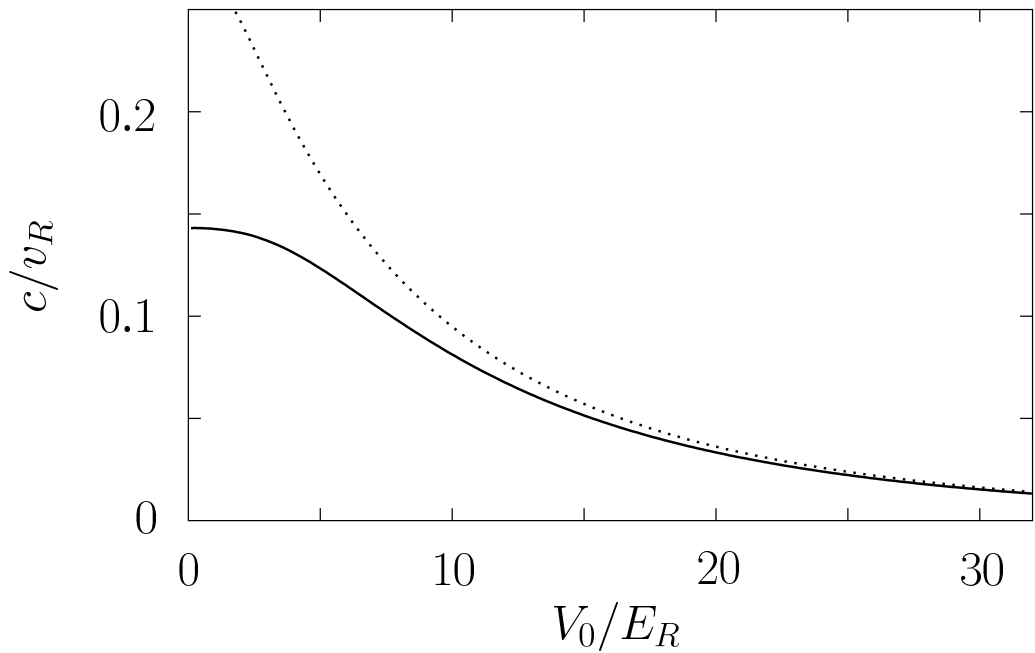} 
\caption{Velocity of sound for a dilute Bose--Einstein condensate 
        with $U_0/\ER = 0.036$ in a 1d optical lattice, as function of  
        the lattice depth. The full line is the result of the Bogoliubov  
        theory, evaluated numerically; the dotted line indicates the  
        asymptotic prediction~(\ref{eq:AS1}).} 
\label{F_1} 
\end{figure}

In order to obtain numerical data for the speed of sound which are  
not restricted to the deep-lattice regime, we calculate the exact Bloch 
waves for the cosine lattice numerically and perform the Bogoliubov  
transformation, assuming the lowest Bloch state $u_{\bm 0}({\bm r})$ to be 
macroscopically occupied; the speed of sound then is extracted directly 
from the slope of the quasiparticle energies $\varepsilon(k)$ for low 
wavenumbers~$k$. Figure~\ref{F_1} shows the result for $d = 1$, as function 
of the lattice depth, for $U_0/\ER = 0.036$, as corresponding, according to  
eq.~(\ref{eq:Uzer}), to the Rubidium data $a = 6$~nm, $\lambda = 852$~nm, 
and $N_s = 1$. Comparison with the deep-lattice formula~(\ref{eq:AS1})  
shows that this estimate describes the numerical data quite well for  
$V_0/\ER > 20$. Here the speed of sound decreases monotonically with  
increasing lattice depth; the perfect agreement of our numerically  
calculated curve with the ones obtained before by  
Kr\"amer {\em et al.\/}~\cite{KraemerEtAl03}, and by  
Menotti {\em et al.\/}~\cite{MenottiEtAl04}, clearly underlines the  
correctness of our reasoning. 
 
%%%%%%%%%%%%%% 
%%%%%%%%%%%% 
 
\begin{figure} 
\includegraphics[width=0.8\linewidth,angle=0]{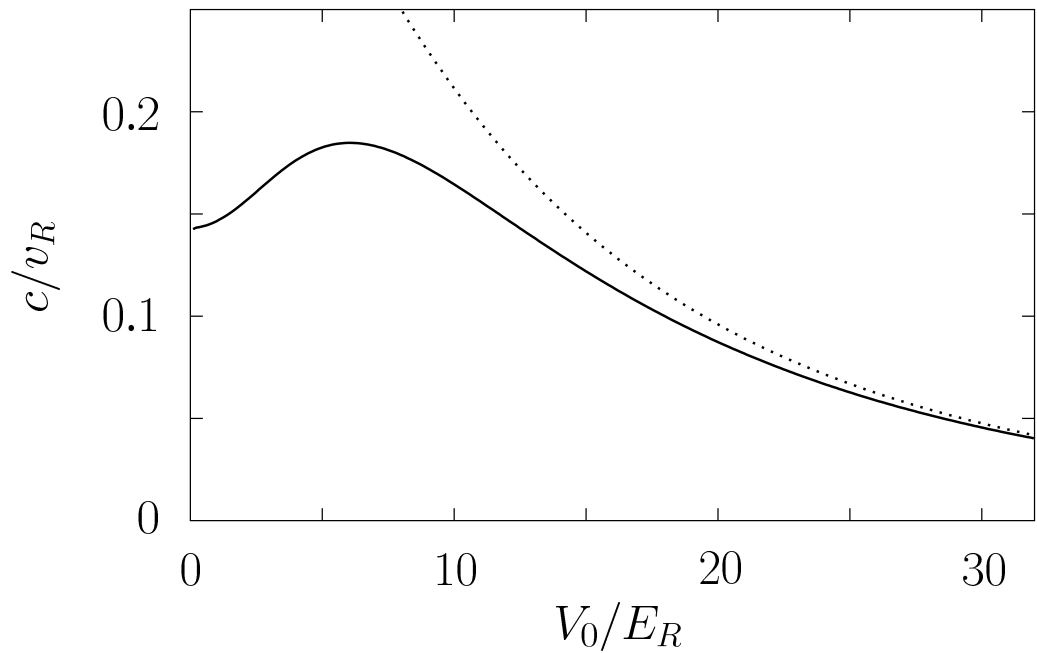} 
\caption{Velocity of sound for a dilute Bose--Einstein condensate 
        with $U_0/\ER = 0.036$ in a 3d optical lattice, as function of  
        the lattice depth. The full line is the result of the Bogoliubov  
        theory, evaluated numerically; the dotted line indicates the 
        asymptotic prediction~(\ref{eq:AS3}). The Mott insulator state,  
        for which the Bogoliubov theory does not apply, occurs for values 
        of $V_0/\ER$ higher than about~$12.3$. The maximum of the velocity 
        of sound lies at $V_0/\ER \approx 6.1$, well in the superfluid  
        regime.} 
\label{F_2} 
\end{figure}

The corresponding data for $d = 3$ are displayed in fig.~\ref{F_2}. 
Now the numerical data for comparatively shallow lattices, for which the  
estimate~(\ref{eq:AS3}) does not apply, show a well-developed maximum 
at $V_0/E_R \approx 6.1$, the speed of sound in a lattice of this depth 
being roughly 30\% higher than in a homogeneous condensate with the same  
average density. Since the superfluid/Mott insulator-transition occurs,  
for the parameters considered, only when $V_0/\ER \approx 12.3$, as  
discussed above, this maximum falls well into the superfluid regime, 
where the simple Bogoliubov approach is still reliable. (For $d = 2$,  
one finds a rather weak maximum at an even lower~$V_0/E_R$.)

%%%%%%%%%%%%%%   
%%%%%%%%%%%% 
 
The appearance of a pronounced maximum of the velocity of sound  
wavepackets reflects the competition between the slowly {\em de\/}creasing  
compressibility~$U^{-1}$, and the exponentially {\em in\/}creasing effective  
mass~$m^*$, with increasing lattice depth. For $d = 1$, the decrease 
of compressibility is so weak that the increasing effective mass wins  
this competition right from the outset, resulting in a monotonically 
decreasing speed of sound. However, for $d = 3$ the stronger decrease 
of compressibility, resulting from the enhanced concentration of the  
condensate at the lattice sites, can over-compensate the increase of $m^*$  
at least in shallow lattices, giving rise to a substantial enhancement of  
the speed of sound before the increasing effective mass again diminishes 
that speed when the lattice is made deeper. An experimental observation 
of this maximum, which should be possible under presently accessible 
laboratory conditions, would constitute an important confirmation of 
our present understanding of the dynamics of Bose--Einstein condensates  
in optical lattices. 
 
\acknowledgments 
This work was supported by the Deutsche Forschungsgemeinschaft through  
the Schwerpunktprogramm SPP~1116, {\em Wechselwirkung in ultrakalten  
Atom- und Molek\"ulgasen\/}.

\end{document}